\documentclass[paper]{aa}

\usepackage{txfonts}

\usepackage{graphicx}

\begin{document}

\def\P{\bar{\Phi}}

\def\st{\sigma_{\rm T}}

\def\vk{v_{\rm K}}

\def\sles{\lower2pt\hbox{$\buildrel {\scriptstyle <}
   \over {\scriptstyle\sim}$}}

\def\sgreat{\lower2pt\hbox{$\buildrel {\scriptstyle >}
   \over {\scriptstyle\sim}$}}

\title{The tension of cosmological magnetic fields 
as a contribution to dark energy}

\author{Ioannis Contopoulos and Spyros Basilakos}
\institute{Research Center for Astronomy, Academy of Athens, 
GR-11527 Athens, Greece, 
\email{icontop@academyofathens.gr; svasil@academyofathens.gr}}

\titlerunning{Cosmological Magnetic Fields}

\date{Received / Accepted }

\abstract
{
We propose that cosmological magnetic fields generated
in regions of finite spatial dimensions may manifest
themselves in the global dynamics of the Universe as
`dark energy'. We test our model in the
context of spatially flat cosmological models by
assuming that the Universe contains 
non-relativistic matter $\rho_m\propto \alpha^{-3}$, 
dark energy $\rho_{Q}\propto \alpha^{-3(1+w)}$,
and an extra fluid with $\rho_{B} \propto \alpha^{n-3}$
that corresponds to the magnetic field. 
We place constraints on the main cosmological parameters 
of our model by combining the recent supernovae type Ia data  
and the differential ages of passively evolving galaxies.
In particular, we find that the model which best reproduces the
observational data when $\Omega_m=0.26$ is one with
$\Omega_{B}\simeq 0.03$, $n\simeq 7.68$, 
$\Omega_{Q}\simeq 0.71$ and $w\simeq -0.8$.

\keywords{Cosmology, magnetic fields}
}

\maketitle

\section{Introduction}
Recent advances in observational cosmology based on the analysis of
a multitude of high quality observational data (type Ia
supernovae, hereafter SNIa, cosmic microwave background, 
large scale structure, age of
globular clusters, high redshift galaxies),
strongly indicate that we are living in a flat
($\Omega=1)$ accelerating Universe 
containing a small baryonic component, non-baryonic
cold dark matter needed to explain the clustering of
extragalactic sources, and an extra component with
negative pressure, usually called `dark energy',
needed to explain the present accelerated expansion of the Universe
(eg. Riess {\em et al.} 1998; Perlmutter {\em et al.}
1999; Efstathiou {\em et al.} 2002; Caldwell 2002; 
Percival {\em et al.} 2002; Spergel {\em et al.} 2003; 
Tonry {\em et al.} 2003; Riess {\em et al.} 2004; 
Tegmark {\em et al.} 2004; 
Corasaniti {\em et al.} 2004).

From a theoretical point of view, various candidates  
for the exotic dark energy have been proposed, 
most of them characterized by an equation of state
$p_{Q}= w\rho_{Q}$ with $w<-1/3$ (see Caldwell 2002; Peebles \& Ratra 2003; 
Corasaniti {\em et al.} 2004 and references therein). 
A particular case of
dark energy is the traditional 
$\Lambda$-model which corresponds to $w=-1$.
Note that a redshift dependence of $w$ 
is also possible but present measurements are 
not precise enough to allow
meaningful constraints (eg. Dicus \& Repko 2004; Wang \& Mukherjee
2006). From the observational point of view and for a flat geometry, 
a variety of studies, and especially the SNIa data, indicate that 
$w< -1$ (eg. Riess {\em et al.} 2004; Basilakos \& Plionis 2005;
Sanchez {\em et al.} 2006; Spergel {\em et al.} 2006; 
 Wood-Vasey {\em et al.} 2007 and references therein). 
For such a fluid the condition $w<-1$ is problematic since
it presents instabilities and causality problems (de la Macorra 2007).
This may be considered as an indication
that the dark energy interacts with another fluid, 
for example magnetic fields.
Tsagas~(2001) gave an interesting
perspective to the problem by claiming that the effect of 
a primordial magnetic
field may resemble that of dark energy through the coupling between
the magnetic field and space time.

In the present paper, we would like to investigate the potential
of present day large scale  magnetic fields to account for the 
effect of dark energy in the observed acceleration of the expansion 
of the Universe. 
As we show in \S~2, if the magnetic field is highly
tangled, it cannot account for the cosmic tension implied by the
presence of dark energy. On the other hand, we argue that, if the cosmic
magnetic field is generated in sources whose overall dimensions
remain unchanged during the expansion of the Universe, 
the stretching of this field by the expansion
generates a tension that behaves
as dark energy. In order to test our model, we
introduce in \S~3 an extra energy density term in the 
cosmological equations which we associate with the magnetic field.
In \S~4 we 
place constraints on the main parameters of our model by
performing a join likelihood analysis utilizing the `gold sample'
of SNIa data (Riess {\em et al.} 2007) as well as the ages
of the passively evolving galaxies (Simon {\em et al.} 2005). 
We conclude with a discussion on the possible values
of cosmological magnetic fields in \S~5.

\section{Tension in a cosmological magnetic field}

It is well known that a positive pressure in the
expanding cosmic fluid contributes to the deceleration
of the expansion. It is easy to understand this when
we realize that every expanding part of the Universe is
pushed against the expansion by its neighboring parts.
As a result, each expanding part performs work against
its surroundings, and thus loses kinetic energy. 
The exact opposite takes
place when the expanding fluid feels a tension force.
In that case, each expanding part is pulled outwards
by its neighbors, and the work done on it by its
neighbors contributes to the acceleration of its
expansion (e.g. Harwit~1982). Dark energy is, therefore,
the cosmic tension that accounts for the observed
present accelerated expansion of the Universe.

When one talks about tension, one immediately comes to think
about magnetic fields. Magnetic field lines may be 
considered as strained ropes, with a highly anisotropic
pattern of tension and pressure. 
We do observe magnetic
fields up to several tens of $\mu G$ in galaxy clusters
(see Carilli \& Taylor~2002 for a review), 
but their origin remains a mystery. 
There is a general belief that cosmic magnetic fields 
are produced through some type of dynamo process 
that amplified a weak protogalactic seed magnetic field of 
the order of $10^{-20}$~G (e.g. \cite{RSS88};
\cite{KCOR97}). In this picture, 
the magnetic field permeates the cosmic
fluid which is assumed highly conductive. In other words,
the sources of the cosmic magnetic field are electric
currents distributed more or less isotropically throughout
the expanding cosmic fluid. 
Furthermore, because of flux
conservation, the magnetic field scales with the
expansion of the Universe as $\alpha^{-2}$,
where $\alpha$ is the Universe scale factor, and therefore,
the magnetic energy contained in any expanding volume
of the Universe scales as $\alpha^{-1}$.
Such a magnetic field generates a positive
isotropic (on average) pressure $p_B=\rho_B/3$, and
$\rho_B\equiv B^2/8\pi \propto \alpha^{-4}$.
The same result may be obtained
after averaging out the magnetic pressure and tension terms.
In other words, the equation of state of a highly
tangled magnetic field is the same as that obtained
for a fluid of highly relativistic particles, and therefore,
it cannot account for the cosmic tension implied by
the presence of dark energy. 
The above led the community to conclude that isotropic tension,
or equivalently negative pressure, is peculiar to a scalar
field (\cite{Z86}).

Here, we would like to investigate the potential of
a different scenario in which the magnetic tension does manifest
itself in the expansion of the Universe as dark energy. 
We thus propose that the sources of cosmological magnetic fields
are of finite dimensions, and that these dimensions remain
unchanged during the expansion of the Universe. 
It is interesting to note that, galaxy clusters are the 
largest gravitationally bound structures in the Universe and as 
such, they have decoupled from the background expansion.
What we have in mind here is some mechanism  that results 
in the generation of magnetic fields of order $B_{0}$
around `sources' of spatial dimensions $r_{0}$ 
(such as galaxies, or clusters of galaxies).
Such a scenario may not be unreasonable.
In fact, we have already proposed a physical picture
where magnetic fields are generated without the need
for a dynamo mechanism, through the
Poynting-Robertson effect on electrons in a highly conducting
plasma around bright gravitating sources such
as active galactic nuclei, black holes, neutron stars, and
protostars (\cite{CK98}; \cite{CKC06}).

If we assume the existence of such sources of cosmic
magnetic fields, it is natural to further assume that
the magnetic field around each source has a dipolar structure.
In that case, the magnetic field drops with distance $r$ as
\begin{equation}
B(r)\approx B_{0}\left(\frac{r}{r_{0}}\right)^{-3}
\label{dipolar}
\end{equation}
in the region influenced by a source of size $r_{0}$ 
at its center. We argue that the Universe may be filled with
several such expanding regions of comoving size $R\alpha(t)$,
which contain  cosmic magnetic sources of size $r_{0}\ll R\alpha$.
The magnetic field energy in each such region is equal to
\begin{equation}
E_B=\int_{r=r_{0}}^{R\alpha} 
\frac{B^2(r)}{8\pi}4\pi r^2 {\rm d}r
\approx \frac{B_{0}^2 r_{0}^3}{6} .
\label{dipoleenergy}
\end{equation}
Obviously, the expansion of the Universe will gradually
stretch each dipolar configuration described by eq.~(\ref{dipolar}) 
into a monopolar configuration described by
\begin{equation}
B(r)\rightarrow B_{0}\left(\frac{r}{r_{0}}\right)^{-2}
\label{monopolar}
\end{equation}
in the region around each source\footnote{
Actually, this is a split-monopole configuration
where space is separated by an equatorial current sheet
discontinuity across which the radial magnetic field
changes direction. Such is the case in stellar magnetospheres.
}.
The magnetic field energy in the expanding region
will approach the value
\begin{equation}
E_B=
\int_{r=r_{0}}^{R\alpha} \frac{B^2(r)}{8\pi}4\pi r^2 {\rm d}r
\rightarrow 
\frac{B_{0}^2 r_{0}^3}{2}\ ,
\label{monopoleenergy}
\end{equation}
which is 3 times larger than the expression in eq.~(\ref{dipoleenergy})!

We showed here that the
magnetic energy $E_B$ contained inside a region of
size $R\alpha(t)$ increases with increasing $\alpha$. 
This may be parametrized
around the present epoch with a simple power law
\begin{equation}
E_B = fB_{0}^2 r_{0}^3 \alpha^n\ ,
\label{energy}
\end{equation}
where, $f\sim 1/3$, and
$n$ is a positive parameter to be determined from
observations (see \S~4). As a result, 
\begin{equation}
\rho_B = \left(\frac{3f}{4\pi}\right)
B_{0}^2 \left(\frac{r_{0}}{R}\right)^3\alpha^{n-3}\ ,\;\; \mbox{and}
\label{energydensity}
\end{equation}
\begin{equation}
p_B = -\frac{n}{3}\rho_B\ .
\label{pressure}
\end{equation}

Note that in our picture, the magnetic field
$B_{0}$ is not primordial, because if that were the case,
by the time the Universe doubles its size,
the magnetic field in each region of magnetic influence
will have effectively completely transformed itself from purely
dipolar to purely (split) monopolar. 
When that happens, the magnetic field
contribution to the cosmological tension dies out.
In our scenario, we expect that the physical mechanism
responsible for the generation of the cosmic magnetic field
(e.g. Contopoulos \& Kazanas~1998)
requires a certain number of years $\tau$ in order
to build a value of the order of $B_{0}$.


\section{Cosmological evolution}

We test our hypothesis by introducing
an extra term that accounts for the magnetic field
in the standard cosmological equations.
In particular, we assume that the
Universe is homogeneous, isotropic, flat, and consists
of the following three components denoted by subscripts `$m$', 
`$Q$' and `$B$' respectively:
non-relativistic matter (with zero pressure), 
an exotic fluid (dark energy),
and a magnetic fluid with an equation of state given
by eq.~(\ref{pressure}). The corresponding
equation of state parameters $w\equiv p_{Q}/\rho_{Q}$ and 
$n\equiv -3p_{B}/\rho_{B}$ are assumed here for simplicity
to be constant. Therefore, 
the evolution of the fluid densities 
$\rho_m$, $\rho_{Q}$ and $\rho_{B}$
is given by $\dot{\rho}_m=-3H\rho_m$, 
$\dot{\rho}_{Q}=-3(w+1)H\rho_{Q}$ and 
$\dot{\rho}_{B}=(n-3)H \rho_{B}$. 
The scale factor of the Universe $\alpha(t)$ evolves according to 
the Friedmann equation:
\begin{equation}\label{eq:11}
H^{2}\equiv \left( \frac{\dot{\alpha}}{\alpha} \right)^{2}=
\frac{8\pi G}{3}(\rho_m+\rho_{Q}+\rho_{B})\ .
\end{equation}
Differentiating the Friedman equation and using at the same time 
the above formalism we obtain
\begin{equation}
\frac{\ddot{\alpha}}{\alpha}=-\frac{4\pi G}{3}[\rho_m+(3w+1)\rho_{Q}+
(-n+1)\rho_{B}] \;\;.
\label{add}
\end{equation}
In this framework, we define the 
density parameters $\Omega_m(\alpha)$, $\Omega_Q(\alpha)$ 
and $\Omega_B(\alpha)$ as
\[
\Omega_m(\alpha)\equiv \frac{\rho_m}{\rho_m+\rho_Q+\rho_B}
\equiv \frac{\Omega_m \alpha^{-3}}{E^2(\alpha)}\ ,
\]
\[
\Omega_Q(\alpha)\equiv \frac{\rho_Q}{\rho_m+\rho_Q+\rho_B}
\equiv \frac{\Omega_Q \alpha^{-3(1+w)}}{E^2(\alpha)}\ ,
\]
\[
\Omega_B(\alpha)\equiv \frac{\rho_B}{\rho_m+\rho_Q+\rho_B}
\equiv \frac{\Omega_B \alpha^{-3+n}}{E^2(\alpha)}\ ,
\]
with
\begin{equation}\label{eq:4}
E(\alpha)=\left(\Omega_m \alpha^{-3}
+\Omega_Q \alpha^{-3(1+w)}
+\Omega_B \alpha^{-3+n}\right)^{1/2}\ .
\end{equation}
Here, the Hubble parameter is given by $H(\alpha)= H_0 E(\alpha)$,
where $H_0$ is the Hubble constant at the present time.
Also, $\Omega_m+\Omega_Q+\Omega_B=1$.

Note that in the context of our model,
$\Lambda$-models correspond to
$(w,\rho_{B})=(-1,0)$ or $(n,\rho_{Q})=(3,0)$, while if $n=0$ or $w=0$,
the extra fluid behaves like pressureless matter.
It is interesting to mention that 
the interplay between the values of 
$w$ and $n$ could yield flat cosmological models 
for which there is not a one-to-one correspondence 
between the global geometry and
the expansion of the Universe. Indeed, in a 
flat low-$\Omega_m$ model with $(w,n)=(-1/3,1)$ 
we have the same dynamics as in an open Universe, despite 
the fact that these models have a spatially flat geometry!

In this cosmological scenario, there is an 
epoch which corresponds to a value of $\alpha=\alpha_{I}$, 
where $\ddot{\alpha_{I}}=0$. This
is called the inflection point.
After that epoch we reach an acceleration phase with
$\ddot{\alpha}>0$.
Eq.~(\ref{add}) thus implies that at the inflection point, 
\begin{equation}
\rho_{m,I}+(3w+1)\rho_{Q,I}+(-n+1)\rho_{B,I}=0 \ ,
\end{equation}
or
\begin{equation}
\Omega_{m}+\Omega_{Q}(3w+1)
\alpha^{-3w}_{I}+
\Omega_{B}(-n+1)\alpha^n_{I}=0 \;\;.
\end{equation}
Therefore, in order for 
the latter equation to contain roots in the interval 
of $\alpha \in [0,1]$, 
we obtain a
theoretical boundary for the possible $(n,w_{Q})$ values, namely
\begin{equation}
3w\Omega_{Q}-n\Omega_{B}<-1\;\;{\rm with}
\;\; w<0\;\;{\rm and}\;\; n>0\ .
\end{equation}
It is obvious that when $\Omega_{B}\rightarrow 0$, the above
constraint tends to the Quintessence case $w<-1/3$, as it should.

\section{Likelihood Analysis}

In order to constrain the cosmological parameters in our model,
we use the `gold' sample of 182 supernovae of
Riess {\em et al.}~2007.
In particular, we define the likelihood estimator\footnote{Likelihoods
are normalized to their maximum values.} as:
${\cal L}^{\rm SNIa}({\bf c})\propto {\rm exp}
[-\chi^{2}_{\rm SNIa}({\bf c})/2]$
with:
\begin{equation}
\chi^{2}_{\rm SNIa}({\bf c})=\sum_{i=1}^{182} \left[ \frac{ {\cal M}^{\rm th}
(z_{i},{\bf c})-{\cal M}^{\rm obs}(z_{i}) }
{\sigma_{i}} \right]^{2} \;\;.
\end{equation}
where ${\cal M}$ is the distance modulus ${\cal M}=5{\rm log}D_{\rm L}+25$,
$D_{\rm L}(z)$ is the luminosity
distance $D_{\rm L}(z)=(1+z)x(z)$, 
$z_{i}$ is the observed redshift, 
$\sigma_{i}$ is the observed uncertainty, and
${\bf c}$ is a vector containing the cosmological 
parameters that we want to fit (Riess {\em et al.}~2007). 
Note, that $x$ is the coordinate distance related to the redshift through 
\begin{equation}
x(z)=\frac{c}{H_{0}} \int_{0}^{z} \frac{{\rm d}y}{E(y)}\;\; .
\end{equation}

\begin{figure}
\includegraphics[angle=0,scale=0.5]{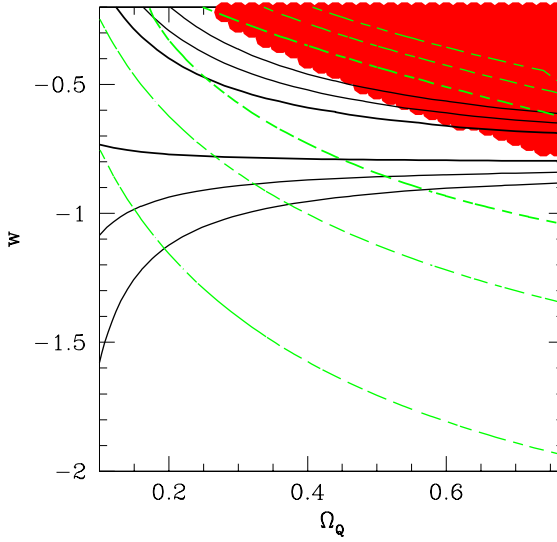}
\caption{Likelihood contours in the $(\Omega_{Q},w)$ plane
for $\Omega_{m}=0.26$. The contours are 
plotted where $-2{\rm ln}{\cal L}/{\cal L}_{\rm max}$ is equal
to 2.30, 6.16 and 11.83, corresponding 
to 1$\sigma$, 2$\sigma$ and 3$\sigma$ confidence level.
Note that the continuous and the dashed lines
correspond to the SNIa and $H(z)$ results respectively,
while the shadowed area is ruled out by stellar ages.
}
\label{fig1}
\end{figure}

We remind the reader that we work here within the framework 
of  flat cosmology ($\Omega_m+\Omega_Q+\Omega_B=1$) with non-zero
large scale magnetic fields $\Omega_{B} \ge 0$.
Furthermore, we use the results of the HST key project 
(Freedman {\em et al.}~2001)
and fix the Hubble parameter to its nominal value 
$H_{0}\simeq 72$ km/s/Mpc. The matter density $\Omega_m$ 
remains the most weakly constrained cosmological parameter. 
In principle, $\Omega_m$ is constrained by the maximum 
likelihood fit  to the WMAP and SNIa data, but in the
spirit of this work, we want to use measures which are completely 
independent of the dark energy component. An estimate of $\Omega_m$ 
without conventional priors is not an easy task in observational cosmology. 
However, many authors using mainly large scale structure studies, 
have tried to put constraints to the $\Omega_m$ parameter. In particular,
from the analysis of the power spectrum, Sanchez {\em et al.}~(2006 and 
references therein) obtain a value
$\Omega_m\simeq 0.24$. Moreover, Feldman {\em et al.}~2003
and Mohayee \& Tully 2005 analyze the peculiar velocity 
field in the local Universe and obtain the values 
$\Omega_m\simeq 0.3$ and $\simeq 0.22$ respectively.
In addition, Andernach {\em et al.}~2005, 
based on the cluster mass-to-light ratio,
claim that $\Omega_m$ lies in the interval $0.15-0.26$ (see
Schindler~2002 for a review). 
In the present paper, we decided to
fix $\Omega_{m}$ to the value $0.26$. 

In this case, the vector of unknown cosmological parameters
is ${\bf c}\equiv (\Omega_{Q},w,n)$. We, therefore,    
sample the various parameters as follows:
the dark energy density $\Omega_{Q} \in [0.01,0.74]$ in steps of
0.01, the dark energy parameter $w\in [-4,-0.1]$ in steps
of 0.02, and the magnetic field scaling parameter $n \in [0.1,10]$ in steps
of 0.02. Doing so, the likelihood function peaks 
at $\Omega_{Q}\simeq 0.5$ ($\Omega_{B}\simeq 0.24$) 
with $w\simeq -1.32$ and $n \simeq 0.44$ (or $w_{B}\simeq -0.15$)
which corresponds to an age of the Universe of 12.3~Gyr.
The latter appears to be
ruled out by stellar ages.  
Indeed, in order to put further constraints on 
our solutions we use additionally
the so called age limit, given by the age of the oldest
globular clusters in our Galaxy ($\simeq 12.5-13$ Gyr;  
Caputo, Castellani \& Quatra 1988; Cayrel {\em et al.}~2001; 
Hansen 2002 and 2004; Krauss~2003 and references therein).
Taking into account the above age limit,
the resulting best fit solution is: 
$\Omega_{B}\simeq 0.04$ ($\Omega_{Q}\simeq 0.7$), 
$w\simeq -0.8$ and $n \simeq 2.52$
(or $w_{B}\simeq -0.84$), corresponding to an age of 13.1~Gyr. 

\begin{figure}
\includegraphics[angle=0,scale=0.5]{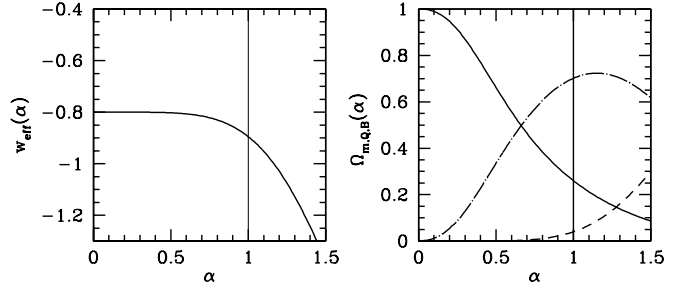}
\caption{Left panel: We show the effective 
equation of state parameter as a function of the scale factor
of the Universe. 
Right panel: We present the evolution of the 
density parameters $\Omega_{m}$ (solid),  
$\Omega_{Q}$ (dot-dashed) and $\Omega_{B}$ (dashed).   
The vertical lines corresponds to the present epoch.
}
\label{fig1}
\end{figure}

In fig.1 (solid lines) we present the 1$\sigma$, 2$\sigma$ and $3\sigma$
confidence levels in the $(\Omega_{Q},w)$ plane by 
marginalizing over $n$. 
It is evident that $\Omega_{Q}$ is
degenerate with respect to $w$
and that all the values in the interval 
$0\le \Omega_{Q} \le 0.74$ are acceptable
within the $1\sigma$ uncertainty. 
Therefore, in order to put further constraints on 
$\Omega_Q$ we additionally use measures of $H(z)$ 
(see Simon {\em et al. 2005}) from the
differential ages of passively
evolving galaxies (hereafter $H(z)$ data). Note that the sample
contains 9 entries. Doing so, the $H(z)$ 
likelihood function can be written as: 
${\cal L}^{H}({\bf c})\propto 
{\rm exp}[-\chi^{2}_{H}({\bf c})/2]$
with:
\begin{equation}
\chi^{2}_{H}({\bf c})=\sum_{i=1}^{9} \left[ \frac{ 
H^{\rm th}(z_{i},{\bf c})-H^{\rm obs}(z_{i})}
{\sigma_{i}} \right]^{2} \;\;,
\end{equation}
where $H(z)$ is the Hubble parameter (see section 3), $H(z)=H_{0}E(z)$. 
The dashed lines in fig. 1 represents 
the $1\sigma$, $2\sigma$, and $3\sigma$,  
confidence levels in the $(\Omega_{Q}, w)$ plane.
In this case, we find that the best fit solution is $\Omega_{Q}\simeq 0.56$
($\Omega_{B}\simeq 0.18$), $w\simeq -0.68$, 
and $n\simeq 8$ (with $T\simeq 13.2$Gyr).   
We now join the likelihoods,
$${\cal L}^{\rm joint}({\bf c})={\cal L}^{SNIa} \times 
{\cal L}^{H}\;\;,$$ and the overall function 
peaks at $\Omega_{Q}=0.71^{+0.03}_{-0.26}$ 
($\Omega_{B}\simeq 0.03)$, $w=-0.80^{+0.14}_{-0.04}$ and 
$n\simeq 7.68^{+2.42}_{-4.00}$ (or $w_B\simeq -2.56$).
Note that, the corresponding age of the Universe is $13.2$Gyr, 
while solving numerically eq. (12) the inflection point is 
at $\alpha_{I}\simeq 0.57$ (or $z_{I} \simeq 0.76$).   
Finally, due to the fact that the association of the 
extra term in the Friedman
equation with the magnetic field is arbitrary,
we may equally well consider the solution
$\Omega_{B}\simeq 0.71$ ($\Omega_{Q}\simeq 0.03$), 
$w\simeq -2.56$ and $n \simeq 2.4$
(or $w_{B}\simeq -0.8$). 

\section{Discussion}
The above values of $\Omega_B$ correspond to an average cosmic
magnetic field 
\begin{equation}
\langle B \rangle
=(8\pi \Omega_{B}\rho_{cr}c^2)^{1/2}\approx 650 \Omega_B^{1/2}
h\;\mu G\ ,
\end{equation}
where, $\rho_{cr}\simeq 1.88\times 
10^{-29}h^{2}\;\mbox{g}\ \mbox{cm}^{-3}$
is the critical density of the Universe; $h$ is the
Hubble constant in units of 100~km/s/Mpc; and
$c$ is the speed of light.
If $\Omega_B\simeq 0.03$, $\langle B \rangle\simeq 80\mu G$,
whereas if
$\Omega_B\simeq 0.71$, $\langle B \rangle\simeq 400\mu G$.
In the present work we do not discuss what physical
mechanism may account for the generation of
such high magnetic fields. 
Magnetic fields of the order of several
tens of $\mu G$ have been observed in the centers of
cooling flow clusters. We refer the reader to
Vogt \& En\ss lin~(2003) (and references therein)
for a detailed discussion of the techniques used to
estimate  such high values of cluster magnetic fields.
Furthermore, a number of authors
have investigated the possibility that $\sim 50\mu G$ fields
may provide magnetic pressure support in cluster
atmospheres (e.g. Loeb \& Mao~1994; Miralda-Escude \& Babul~1995;
Dolag \& Schindler~2000; but see also
Rudnick \& Blundell~2003). We argued that dark energy
(or equivalently cosmic tension) manifests itself 
as a tension agent between galaxy clusters,
and as such, it obviously acts in
intra-cluster space. Moreover, there
is a theoretical indication that
strong magnetic fields lie in regions
of significantly reduced plasma density
(e.g. Gazzola {\em et al.}~2007), which would make
observations of cosmic magnetic fields through 
Farady rotation or X-ray emission in
intra-cluster cosmic voids very difficult.
We must keep in mind that the study of
cosmic magnetic fields on scales of clusters of galaxies
is a fairly new area of research (see Carilli \& Taylor~2002
for a review), and therefore, we cannot preclude
future observational surprises.
In particular, 
any observation of magnetic fields of the order of $100\mu G$
over cosmological scales would give credence to our scenario. 

We would like to end this section with a short
discussion on the equation of state of the dark energy.
As we mentioned in the introduction, there is strong indication
for an equation of state more complicated than the
simple assumption of a constant ratio between the pressure
and the energy density. We may thus combine our two
dark energy fluids into one with an effective
dark energy parameter
\begin{equation}
w_{\rm eff}=\frac{P_{\rm eff} }{\rho_{\rm eff}}=\frac{P_{Q}+P_{B}}
{\rho_{Q}+\rho_{B} } \;\;.
\end{equation}
Using the evolution of $\rho_{Q}$ and $\rho_{B}$
the effective dark energy parameter as a function of time is given by
\begin{equation}
w_{\rm eff}(\alpha)=\frac{3w-n\nu \alpha^{3w+n}}{3(1+\nu \alpha^{3w+n})} 
\;\;\;\;{\rm where}\;\;\;\;\nu=\frac{\Omega_{B}}{\Omega_{Q}} \;\;.
\end{equation}
Using our best fit parameters, 
$w_{\rm eff}\sim -0.9$ for $\alpha\sim 1$, 
$w_{\rm eff}\sim -0.8$  in the limit $\alpha \ll 1$,
and $w_{\rm eff}\sim -2.6$  in the limit $\alpha \gg 1$.
In the left panel of fig.~2, the solid line shows
the evolution of the effective equation of state parameter
as a function of the Universe scale factor.
A first order Taylor expansion around the present epoch 
(Linder~2003) yields
\begin{equation}
w_{\rm eff}(\alpha)\simeq -0.87+0.36(1-\alpha) \;\;.
\end{equation}

In the right panel of fig.~2 we show the evolution of the density 
parameters $\Omega_{m}$ (solid line),  
$\Omega_{Q}$ (dot-dashed) and $\Omega_{B}$ (dashed). 
It is interesting that, although at the present time the
dark energy in dominant, before 
the inflection point ($\alpha_{I} \simeq 0.57$) the Universe
was matter dominated, i.e. $\Omega_m(\alpha\ll 1)\approx 1$.
In fact, we estimate that prior to the inflection point,
$\Omega_{B}\leq 0.2\%\Omega_m$, which corresponds to
an average cosmic magnetic field
$\langle B \rangle\leq 30\mu G$. This value 
was well under equipartition, and therefore, 
one may argue that, at an early enough epoch, 
matter is able to generate the cosmological magnetic fields
required in our scenario (for example, through the 
Poynting-Robertson mechanism described 
in Contopoulos \& Kazanas~1998). 

We conclude with a summary of the main elements of our scenario:
\begin{enumerate}
\item The cosmological magnetic field is generated in
sources of characteristic size $r_{0}$ with characteristic
value $B_{0}$. The main idea is
that the size of these sources does not follow the overall
expansion of the Universe. We know that the expansion of the
Universe manifests itself over length scales larger than
the typical size of clusters of galaxies.
This leads us to suggest that the size of our putative
magnetic field sources is of the order of a few Mpc.
\item Each source is associated with a region of magnetic
influence around it where the large scale field
is due to the central source. 
The sources are uniformly and isotropically distributed
throughout the Universe.
\item As the Universe expands, the magnetic field in
each region of influence is stretched, and
the total magnetic field energy grows.
This results in the acceleration of the expansion.
The acceleration will decrease unless
new sources are continuously generated throughout the Universe.
\item We model the effect of the magnetic field with
an extra term $\rho_B\propto \alpha^{n-3}$ in the
Freedman equations. The model that best reproduces the
observational data when $\Omega_m=0.26$ is one with
$\Omega_{B}\simeq 0.03$, $n\simeq 7.68$, 
$\Omega_{Q}\simeq 0.71$ and $w\simeq -0.8$,
which yields an average cosmic magnetic field of
the order of $\sim 80\mu G$. 
Obviously, we may equally well consider
the solution $\Omega_{B}\simeq 0.71$, $n\simeq 2.4$, 
$\Omega_{Q}\simeq 0.03$ and $w\simeq -2.56$.
The latter corresponds to an average cosmic
magnetic field of the order of $\sim 400\mu G$.
\end{enumerate}



\begin{thebibliography}{}

\bibitem[]{}Andernach, H., Plionis, M., Lopez-Cruz, O., Tago, E., 
Basilakos, S., 2005, Astronomical Society of the Pacific 
Conference Series, vol. 329, p. 289-293. Nearby Large-Scale 
Structures and the Zone of Avoidance, Proceedings of a meeting 
held in Cape Town, South Africa, March 28 -- April 2, 2004, San 
Francisco: Astronomical Society of the Pacific, 2005.
\bibitem[]{} Basilakos, S. \& Plionis, M., 2005, MNRAS, 360, L35
\bibitem[]{}Caldwell, R. R., 2002, Physics Letters B, 545, 23
\bibitem[]{}Caputo, F., Castellani, V., Quarta, M. L, 1988, 
A Self-Consistent Approach to the Age of Globular Cluster M15,
The Early Universe: Reprints Edited by Edward W. Kolb and 
Michael S. Turner. Frontiers in Physics, Reading: Addison-Wesley, p.263
\bibitem[]{}Carilli, C. L. \& Taylor, G. B. 2002, Ann.Rev.A\& A, 40, 319
\bibitem[]{} Cayrel, R., {\em et al.} 2001, Nature, 409, 691
\bibitem[Contopoulos \& Kazanas 1998]{CK98}
Contopoulos, I. \& Kazanas, D. 1998, ApJ, 508, 859
\bibitem[Contopoulos, Kazanas \& Christodoulou 2006]{CKC06}
Contopoulos, I., Kazanas, D. \& Christodoulou, D. M. 2006, 
ApJ, 652, 1451
\bibitem[]{}Corasaniti, P. S., 
Kunz, M., Parkinson, D., Copeland, E. J., Bassett,
B. A., 2004, Phys. Rev. Lett., 80, 3006
\bibitem[]{} de la Macorra, A., 2007, submitted (astro-ph/0701635)
\bibitem[]{}Dicus, D. A. \& Repko, W.W., 2004, Phys.Rev.D, 70, 3527, 
\bibitem[]{}Dolag, K. \& Schindler, S. 2000, A\& A, 364, 491
\bibitem[]{} Efstathiou, G. {\em et al.} 2002, MNRAS, 330, L29
\bibitem[]{}Feldman, H. {\em et al.} 2003, ApJL, 596, L131
\bibitem[]{}Freedman, W., L., {\em et al.}, 2001, ApJ, 553, 47
\bibitem[]{}Gazzola, L., King, E. J., Pearce, F. R.
\& Coles, P. 2007, MNRAS, 375, 657
\bibitem[]{}Hansen B. {\em et al.} ApJL, 2002, 574, 155
\bibitem[]{}Hansen B. {\em et al.} ApJS, 2004, 155, 551
\bibitem[Harwit 1982]{H82}
Harwit, M. 1982, in Astrophysical Concepts (Wiley: New York)
\bibitem[]{}Krauss, L. M., 2003, ApJ, 596, L1
\bibitem[Kulsrud {\em et al.} 1997]{KCOR97}
Kulsrud, R. M., Cen R., Ostriker, J. P. \&
Ryu, D. 1997, ApJ, 480, 481
\bibitem[]{}Linder E. V., Phys. Rev. Lett., 2003, 90, 1301
\bibitem[]{}Loeb, A. \& Mao, S. 1994, ApJ, 435, 109
\bibitem[]{}Miralda-Escude, J. \& Babul, A. 1995, ApJ, 449, 18
\bibitem[]{}Mohayaee, R. \& Tully, B., 2005, AJ, 130, 1502
\bibitem[]{}Peebles P. J. E., \& Ratra, B., 2003, RvMP, 75, 559
\bibitem[]{}Perlmutter, S. {\em et al.} 1999, ApJ, 517, 565
\bibitem[]{}Percival, J. W. {\em et al.} 2002, MNRAS,337, 1068
\bibitem[]{} Riess, A. G. {\em et al.} 1998, AJ, 116, 1009
\bibitem[]{} Riess, A. G. {\em et al.} 2004, ApJ, 607, 665
\bibitem[]{} Riess, A. G. {\em et al.} 2007, ApJ, 659, 98
\bibitem[]{}Rudnick, L. \& Blundell, K. M. 2003, ApJ, 588, 143
\bibitem[Ruzmaikin, Shukurov \& Sokoloff 1988]{RSS88}
Ruzmaikin, A. A., Shukurov, A. M. \& Sokoloff, D. D. 1988,
in Astrophysics and Space Science Library, Magnetic Fields
in Galaxies (Dordrecht: Kluwer)
\bibitem[]{}Sanchez, A. G., Baugh, C. M., Percival, W. J., Peacock,
  J. A., Padilla, N. D., Cole, S., Frenk, C. S., Norberg, P., 2006, MNRAS,
  366, 189
\bibitem[]{}Schindler S., Space Science Reviews, 2002, 100, 299
\bibitem[]{}Simon, J., Verde, L., Jimenez, R., 2005, Phys. Rev. D., 
71, 123001
\bibitem[]{}Spergel, D. N. {\em et al.} 2003, ApJS, 148, 175
\bibitem[]{}Spergel D. N. {\em et al.} ApJ, 2007, in press
(astro-ph/0603449)
\bibitem[]{}Tegmark M. {\em et al.} 2004, Phys.Rev.D., 69, 3501
\bibitem[]{}Tonry {\em et al.} 2003, ApJ, 594, 1
\bibitem[]{}Tsagas, C., 2001, Phys. Rev. Letters, 86, 5421
\bibitem[]{}Vogt, C. \& En\ss lin, T. A. 2003, A\& A, 412, 373
\bibitem[]{}Wang, Y. \& Mukherjee, P., 2006, ApJ, 650, 1
\bibitem[]{}Wood-Vasey, M. W. {\em et al.} 2007, submitted (astro-ph/0701041)
\bibitem[Zeldovich 1986]{Z86}
Zeldovich, Y. B. 1986, Sov. Sci. Rev. E Astrophys. Space Phys, 5, 1

\end{thebibliography}
\end{document}